\documentclass[
 preprint,
 superscriptaddress,
 amsmath,amssymb,
 aps,
 pra,
 floatfix
]{revtex4-1}
\usepackage{xcolor}
\usepackage{graphicx}
\usepackage{bm}
\usepackage{floatrow}

\begin{document}

\title{Mechanical prions: Self-assembling microstructures}%

\author{Mathieu Ouellet\footnote{Except for the first author, all authors are listed alphabetically. \\ Corresponding author: ouellet@seas.upenn.edu.}}
\affiliation{
Department of Electrical \& Systems Engineering, University of Pennsylvania, Philadelphia, PA 19104 USA
}

\author{Dani S. Bassett} 
\affiliation{
Department of Bioengineering, University of Pennsylvania, Philadelphia, PA 19104 USA
}
\affiliation{
Department of Physics and Astronomy, University of Pennsylvania, Philadelphia, PA 19104 USA
}
\affiliation{
Department of Electrical \& Systems Engineering, University of Pennsylvania, Philadelphia, PA 19104 USA
}
\affiliation{
Department of Neurology, University of Pennsylvania, Philadelphia, PA 19104 USA
}
\affiliation{
Department of Psychiatry, University of Pennsylvania, Philadelphia, PA 19104 USA
}
\affiliation{
Santa Fe Institute, Santa Fe, NM 87501 USA
}
\affiliation{
Montreal Neurological Institute, McGill University, Montreal, QC H3A 2B4, Canada
}

\author{Lee C. Bassett}
\affiliation{
Department of Electrical \& Systems Engineering, University of Pennsylvania, Philadelphia, PA 19104 USA
}

\author{Kieran A. Murphy}
\affiliation{
Department of Bioengineering, University of Pennsylvania, Philadelphia, PA 19104 USA
}

\author{Shubhankar P. Patankar}
\affiliation{
Department of Bioengineering, University of Pennsylvania, Philadelphia, PA 19104 USA
}

\date{\today\newpage}

\begin{abstract}
Prions are misfolded proteins that transmit their structural arrangement to neighboring proteins.
In biological systems, prion dynamics produce complex functional outcomes, ranging from useful long-term memory information to harmful spongiform encephalopathies.
Yet, an understanding of prionic causes has been hampered by the fact that few computational models exist that allow for experimental design, hypothesis testing, and control.
Here, we identify essential prionic properties and present a biologically inspired model of prions using simple mechanical structures capable of undergoing complex conformational change. 
We demonstrate the utility of our approach by designing a prototypical mechanical prion and validating its properties experimentally.
Our work provides a design framework for harnessing and manipulating prionic properties in natural and artificial systems. 
\end{abstract}

\maketitle

\newpage
\section{Introduction}
Prions are shape-shifting proteins notorious for their ability to cause deadly transmissible diseases \cite{aguzzi2012prion,prusiner1998prions}.
These include scrapie in sheep and goats, bovine spongiform encephalopathy in cattle, and Creutzfeld-Jakob disease in humans.
Despite such examples of malignancy, prionic behavior can also be advantageous in specific contexts. 
Proteins with prionic appendages facilitate the formation and maintenance of long-term memories, while fungi such as yeasts rely on prions to transmit heritable characteristics \cite{wickner2015yeast}.
Given these dual roles, prions are essential targets for investigation, both for understanding disease and exploring potential applications.

From an engineering standpoint, the utility of prion proteins has long been recognized.
Prionic aggregates are generally highly resistant to proteolytic digestion, heat, chemical denaturation, and solubility in non-ionic detergents \cite{diaz2018prion}.
One standard method for integrating prionic properties into nanosystems is constructing proteins with a known prion domain \cite{Colby2011novo}.
This approach has been used to build nanowires \cite{scheibel2003conducting,men2010auto}, biofilms \cite{altamura2017synthetic}, dynamic hydrogels \cite{das2016implantable}, proton-conducting biomaterials \cite{navarro2024harnessing, li2012biodegradable} ,
 pharmacological agents \cite{kuang2014prion,wang2020multifunctional}, and biosensors \cite{diaz2021functionalized,behbahanipour2024bioengineered,behbahanipour2023oligobinders,men2010auto}.
These efforts demonstrate the utility of prion-inspired mechanisms far beyond their origins in biology.

\begin{figure}[h!]
\includegraphics[width=0.8\textwidth]{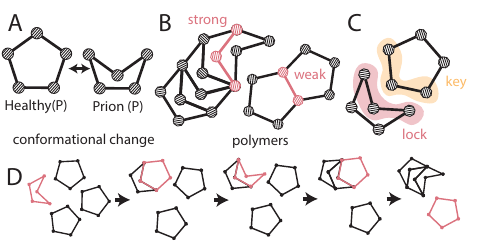}%
\caption{\label{fig:fig0}  \textbf{Essential prionic properties.} (\textbf{A}) The system can exist in two distinct states, namely the prion (P) and healthy (H) conformations, respectively. (\textbf{B}) The prion conformation can form polymers with other prions, whereas the healthy conformation cannot. (\textbf{C}) A lock-and-key mechanism involving a prion conformation and a healthy conformation enables the formation of dimers. (\textbf{D}) Conformational changes propagate infectiously in a monodisperse solution when a single prion is present.}
\end{figure}

At its core, all prionic behavior---whether biological or synthetic---is driven by the infectious propagation of conformational changes \cite{slepoy2001statistical,weickenmeier2018multiphysics}.
Additionally, prion proteins must possess several fundamental properties.
First, they must have at least two stable conformations.
For consistency with biological prions, we refer to these conformations as healthy (H) and prionic (P) (Figure \ref{fig:fig0}A). 
Second, the P conformation must be able to form stable polymers, such that P-P dimers are more stable than H-H dimers (Figure \ref{fig:fig0}B).
Third, the H conformation must be able to interact with the P conformation to form an unstable H-P dimer. 
The H-P dimer, in turn, must be able to convert to a P-P dimer through a reaction that cannot occur solely with monomers (Figure \ref{fig:fig0}C).
If these minimal conditions are met, we posit that a propagation mechanism similar to that occurring in biological prions can be replicated in a system of simple mechanical elements that we call \emph{mechanical prions} (Figure \ref{fig:fig0}D).

Physical studies of biological prions have primarily taken a macroscale, statistical view, focusing on chemical rate modeling \cite{ferreira2003stochastic,michaels2016hamiltonian}. 
However, a physical model reproducing the critical elements of prions---including their ability to exist in distinct conformations, self-assemble into the prion state, and self-replicate---has yet to be developed and examined.
This work addresses this gap by presenting a biologically-inspired mechanical model of prions.
We begin with a mechanical bar-joint linkage model.
Next, we search the space of resulting structures to identify candidate prions and validate the presence of prionic properties.
Finally, we construct a tabletop macroscale mechanical prion with the minimal properties required to elicit prionic behavior.
Our work presents a principled approach to designing and building mechanical structures that exhibit prionic behavior.

\section{Mechanical Prions}

\textbf{A bar-joint linkage model of prions.} Inspired by prior work on conformational changes \cite{rocklin2018folding, Stern2018, kim2019conformational}, we employ a bar-joint linkage model \cite{chen2022modular} within a thermal bath \cite{mannattil2022thermal}.
Each structure comprises $N$ nodes, divided into two groups: exterior nodes, which enable polymer binding and interior nodes, for structural stability.
Interactions between linked nodes are modeled using a harmonic potential common for coarse-grained modeling of proteins \cite{procyk2021coarse,rocklin2018folding} and a Lennard-Jones (LJ) 12-6 potential to allow for interactions between nodes across prions \cite{wang2020lennard}.
We present results in LJ units, where the LJ energy minimum, $\varepsilon$, is set to $0.05K \ell^2$, where $K$ is the harmonic spring constant and $\ell$ is the length of the external bonds.
We set the LJ length scale, $\sigma$, such that the energy minimum between two nodes occurs at a separation of $0.05\ell$.
To incorporate thermal fluctuations, we model system dynamics using the Langevin formalism in the Large-scale Atomic/Molecular Massively Parallel Simulator (LAMMPS) \cite{dias2021molecular, supp}.

\begin{figure}
\includegraphics[width=0.6\textwidth]{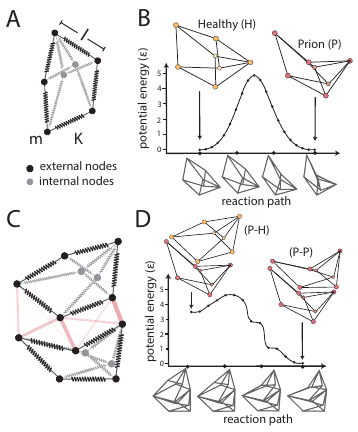}%
\caption{\label{fig:fig1} \textbf{Bar-joint mechanical prion model.} (\textbf{A}) Schematic of the structure.
(\textbf{B}) Bond energy of the free system evaluated along the reaction path between the H and P conformations. (\textbf{C}) Schematic of the P and H conformations interacting.
(\textbf{D}) Potential energy of the bond between the P-H dimer and the P-P dimer evaluated along the reaction path.}
\end{figure}

\textbf{Designing mechanical prions.} When searching the space of bar-joint linkages for viable prions, several design choices are pertinent: the exterior shape for stable configurations (defining a possibly non-convex polygon), the number and position of the internal nodes, and the number and lengths of the edges connecting the internal and external nodes. 
We constrain our search for mechanical prions to the space of regular pentagons, also described as a $4$-torus \cite{shimamoto2005spaces}.  
With the number of external nodes set to five, we aim to find two distinct configurations representing the H and P structures.
First, we require that the P structure self-bind with lower energy than the H structure.
We search for a P polygon where three external nodes can align with three nodes of another P polygon (Figure \ref{fig:fig0}B).
Second, the binding process must allow the formation of a tower-like structure of P polygons, while the H polygon should resist self-binding.
To ensure this is the case, we limit the maximum number of interacting nodes between two H polygons to two (Figure \ref{fig:fig0}B).
Third, we search for a lock-and-key mechanism \cite{woods2022shape}, whereby the P and H polygons interact via three external nodes, with the H polygon requiring a small deformation for binding to the P polygon (Figure \ref{fig:fig0}C). 
This imperfection disrupts the rate symmetry between free and bonded states (Figure \ref{fig:fig0}D).

To guarantee mechanical rigidity, the external nodes defining the polygons must connect to a set of internal nodes \cite{chen2022modular} (Figure \ref{fig:fig1}A), thereby allowing for two stable conformations of a single structure. 
These conformations have identical energy minima (Figure \ref{fig:fig1}B) \cite{kim2019conformational,kim2022nonlinear}.
The two conformations are isolated by an activation energy barrier, which prevents transitions from one state to another. 
The two interacting structures exhibit prominent interactions within the lock-and-key region, whereas other interaction configurations are inhibited by design (Figure \ref{fig:fig1}C).
The potential energy in the P-P configuration is an energy minimum and, at first order, equals that of two separate P structures. 
By contrast, the potential energy of the P-H configurations is notably higher due to misaligned lock-and-key areas (Figure \ref{fig:fig1}D). 
The potential energy of the barrier remains largely unaffected.
This reduced activation energy for P-H to P-P transitions alters the symmetry of configuration rates \cite{supp}, driving the prionic effect.

This search reveals several candidate mechanical prions in the 4-torus space of pentagonal bar-joint linkages\footnote{For ease of computation, we self-constrain the design process to the space of pentagons; however, our method can be applied more generally.}.
Although their static energy landscape suggests potential prionic behavior, evaluating the candidate structures under dynamic conditions remains essential.
Multiple undesired effects can appear in a dynamic regime, preventing crucial transitions from P-H to P-P. 
Examples of these effects include internal node binding, extra undesirable stable configurations, and inhibition of the forward reaction due to entropy \cite{tzeng2012protein}. 
The initial search phase does not address these challenges directly. 
To address these concerns, we explicitly evaluate candidate pairs of configurations under Langevin dynamics.

\begin{figure}
\includegraphics[width=0.55\textwidth]{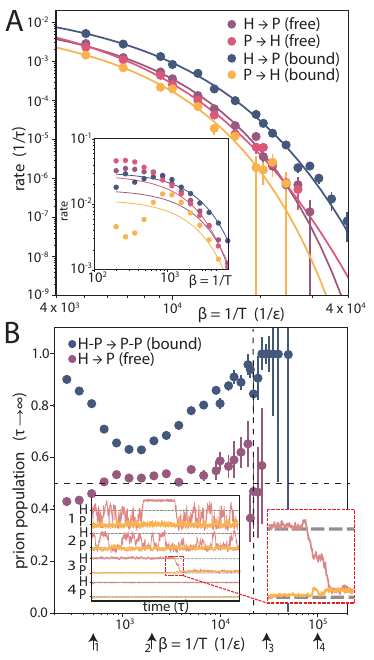}%
\caption{\label{fig:fig2} \textbf{Dynamical properties.} (\textbf{A}) The transition rate between the two conformations (P for prion and H for healthy) in the free and bound cases at low temperatures fit with a super-Arrhenius form. The inset shows the deviation from the fitting at higher temperatures. (\textbf{B}) The two curves show the equilibrium distribution of the prion population in the free (in dark purple) and bound (in dark blue) cases given the inverse temperature. The inset shows four examples at various temperatures of the dynamics of each monomer (P and H) inside a dimer (PH). The temperatures are indicated by arrows beneath the plot. Here, we used the same reaction coordinates as those in Figure 1.C. }
\end{figure}

\textbf{Kinetics of forward and backward reactions are assymetric.} We measure the sum of squared differences between the internal angles of dynamically evolving structures and those in static reference conformations corresponding to the healthy or prionic states.
We assign each structure the label of its closest static counterpart.
We use these labels to analyze Langevin dynamics via a discrete-time Markov model, deriving transition rates through model fitting over a coarse-grained time-scale \cite{supp}.
We investigate the evolution of an ensemble comprising P and H structures and track the concentrations of both interacting (bound P-H and P-P) and non-interacting (free P and H) components. 
Reaction rates for the candidate pentagonal structure are plotted in Figure \ref{fig:fig2}A.
In the free scenario, both the H$\to$P and P$\to$H rates are indistinguishable within margins of uncertainty.
However, in the bound scenario, the H$\to$P rate is up to two orders of magnitude faster, breaking the symmetry between the forward and backward reactions. 
At lower temperatures ($\beta\varepsilon=\frac{\varepsilon}{k_\mathrm{B}T} > 3 \times 10^4$), P$\to$H conversions in the bound state are infrequent; none are observed within a time scale feasible for simulation.
These reaction rates indicate that the designed structure exhibits prionic behavior.

\textbf{Rate of prion propagation is unexplained by the Arrhenius model.} Equilibrium distributions derived from a Markov model based on observed transition rates are shown in Figure \ref{fig:fig2}B.
As expected for a system containing prions, the stationary distribution is asymmetric; bound systems approach 100\% prion conversion, whereas free systems approach only 50\% prion conversion. 
At low temperatures ($\beta\varepsilon > 10^{3}$), both free configurations are thermodynamically stable, and no other conformations are present.
We arrive at equilibrium distributions (Figure \ref{fig:fig2}B) using simulation rates derived from an extended Arrhenius model \cite{kohout2021modified}. 
The model accurately characterizes dynamic behavior at low temperatures (Figure \ref{fig:fig2}A).
However, we observe non-monotonic fluctuations (U-shaped) for the bound configuration at high temperatures that cannot be accurately described by an Arrhenius-type equation (Figure \ref{fig:fig2}A, inset). 
This phenomenon is potentially explained by the ability of the bound networks to attain new, slightly deformed conformations at high temperatures.
We also observe super-Arrhenius behavior, or lower-than-expected reaction rates, at low temperatures.
This effect also seen in the kinetics of protein folding \cite{mallamace2016energy}, in the collective behavior of glassy systems \cite{zhao2013using, jaiswal2016correlation}, and in enzyme-catalyzed reactions \cite{carvalho2016description}, is generally related to an increase in the activation energy or a change in the transition state at lower temperatures \cite{aquilanti2017kinetics}. 

\textbf{Interaction between P and H monomers alters the dynamics of bound and unbound dimers.}
In the inset of Figure \ref{fig:fig2}B, we display time traces that illustrate the conformational changes of the bound structure, all initiated from the P-H conformation.
New stable conformations are reached at high temperatures, as seen in the first time trace (Figure \ref{fig:fig2}B), where the curves exceed the H conformation and remain stable for an extended period before moving to the P conformation.
The third trace displays the system at a temperature where unbound transitions are exceedingly rare.
In this trace, bound H visibly transitions to the P conformation, as highlighted in the magnified portion of the inset.
We observe a subtle conformational shift in the P structure as the H transforms, appearing to begin when the H structure is halfway through its transition.
This timing suggests that the P structure (to which the H binds) actively participates in the conversion process.
Furthermore, the final stability of P-P dimers is not simply an outcome of the stability of individual P and H monomers.
This stability of large assemblies in the prionic state is crucial for prion functionality.
We note a difference in the levels of random fluctuations between the healthy and prion states at all temperatures (see Figure \ref{fig:fig2}A inset).
These differences are likely due to the empty key-lock site for the healthy configuration, making structures less rigid.

\begin{figure}
\includegraphics[width=0.7\textwidth]{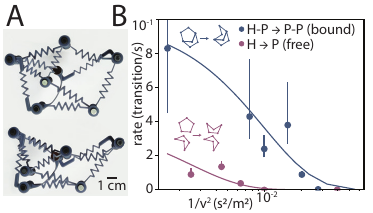}%
\caption{\label{fig:fig3} \textbf{External mechanism and macroscopic model.} (\textbf{A}) (top) Macroscopic model exhibiting the healthy conformation. (bottom) Macroscopic model exhibiting the prion conformation. (\textbf{B}) The transition rate exhibited by the macroscopic model is similar to the transition rate exhibited by simulated structures. 
The inset shows one of the macroscopic models used in the experiment. Further details can be found in the Supplement.}
\end{figure}

\textbf{The lock-and-key mechanism is crucial for eliciting prionic behavior.} 
Intuitively, the lock and the key regions should have the same geometry to enable the alignment of the lock with the key.
However, when two networks with perfect alignment bind to one another, the overall shape of the constituent networks does not change.
This rigidity is due to Lennard-Jones potentials existing at significant levels only between matched node pairs and not between distant misaligned nodes.
Therefore, the lock-key areas must necessarily be misaligned to force internal reorganization in the H monomer.
This misalignment of external nodes breaks the symmetry in the rates between the free and bonded states (see Supplement). 
While the external node configurations establish the lock-and-key mechanism, the positioning of internal nodes also plays a crucial role.
Internal nodes ensure the stability of the mechanical prion while simultaneously facilitating conformational changes.
Their placement greatly influences the transition temperature by changing the structural stability. 
We find that prionic behaviors of mechanical structures are robust to a wide range of choices of the precise locations of the internal nodes \cite{supp}.

\textbf{Tabletop prion model.} We show that a bar-joint linkage model can be successfully coupled with a careful search process to uncover mechanical structures capable of undergoing prion-like conformational changes.
To experimentally validate this finding and demonstrate the practical utility of our approach, we build a tabletop environment where the dynamics of the designed structural configurations can be evaluated directly in a proof-of-concept experiment.
We 3D-print spring-like edges that can extend and contract (Figure \ref{fig:fig3}B) and place magnets at each node to allow interactions across structures.
We simulate a thermal bath using a stepper motor to agitate the structures at varying speeds.
Transition rates observed in the free and bound states qualitatively align with our theoretical predictions, showing the expected rate asymmetry in the forward and backward transitions (Figure \ref{fig:fig3}B).
In the supplementaty materials, we include videos of the P-H dimer converting to the P-P configuration and additional details of the macroscale model.

\section{Discussion}

This study introduces a biologically inspired mechanical model of prions, leveraging simple bar-joint linkage systems to emulate prionic behaviors. 
Through computational simulations and experimental validation, we demonstrate that our model successfully replicates the essential characteristics of biological prions, including asymmetric reaction rates and aggregation.
These findings provide a foundational framework for exploring prion-like dynamics in engineered systems, opening new pathways for applications ranging from self-assembling metamaterials to molecular-scale sensing. 

\begin{figure}
\includegraphics[width=0.75\textwidth]{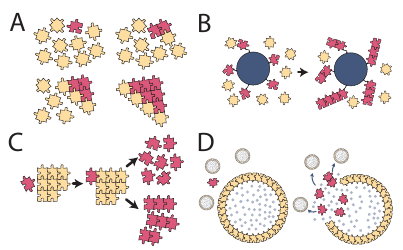}
\caption{\label{fig:fig4} \textbf{Potential applications.} (\textbf{A}) Prions could have multiple binding sites, thereby allowing for higher dimensional structure. Here, a simple 2D sheet is represented. (\textbf{B}) Prions could be utilized in the functionalization of nanoparticles, where a change in conformation could result in a change in solubility or the formation of structures such as fibrils. (\textbf{C}) Prions could be used to create structures that are capable of disassembling themselves, triggered by a change in the conformation of one of their components. (\textbf{D}) The transformation of healthy assembled structures into non-interacting prions can be leveraged as a switching mechanism through the disassembly process.}
\end{figure}

In aggregate form, mechanical prions offer a variety of valuable architectures, powerful functions, and capacities for design and control.
Prion-like mechanical networks can help build fast and irreversible sensors, leveraging their unique polymerization properties and switching capabilities.
The switching rates of these structures can be fine-tuned by designing their binding sites to promote aggregation over fragmentation preferentially, with a specific focus on the large-scale structures that are targeted for production \cite{tanaka2006physical, matveev2017oscillations}.
Similar to their biological counterparts that produce amyloid plaques, mechanical prions can assemble into durable 2D sheets (Figure \ref{fig:fig4}A) \cite{diaz2018prion} and 1D fibrils \cite{lee2015self,levin2020biomimetic,zhang2015unfolding} (Figure \ref{fig:fig4}.B).
Such constructions have been explored, for example, using prion-inspired peptides \cite{di2020engineered, diaz2018minimalist, navarro2023amyloid, barbee2021protein, jankovic2021exploiting}. 
Monodisperse prion solutions, which restructure after seeding, have potential utility in scavenging and control processes \cite{navarro2023amyloid}.
Recent advancements in DNA origami have enabled the fabrication of miniature mechanical structures with joints, sliders, and hinges, indicating the potential for constructing mechanical prions using similar methods \cite{dey2021dna, wang2023dynamic, huang2021integrated, modi2009dna, nummelin2020robotic}.

Prions are often associated with pathological aggregates, but recent studies have also outlined their role in various beneficial biological functions \cite{Levkovich2021microbial, kuang2014prion}.
Not all aggregates are pathological, with some having potential functional utility.
For instance, ``healthy'' configurations can be aggregated to design self-disassembling scaffolds (Figure \ref{fig:fig4}C).
Inspired by these findings, future approaches might introduce prions into synthetic materials, potentially for drug delivery systems where a prion-composed shell safely disintegrates at its intended target \cite{verma2013self} (Figure \ref{fig:fig4}D).

Prions are excellent candidates for use in engineered microscale systems.
They are stable in two distinct conformations and can link with other prions to form supramolecular lattice-like aggregates.
In a self-perpetuating process, when a prion protein undergoes a conformational change to the prionic state, it induces its neighbors to undergo the same transformation.
These properties of stability and self-propagation enable the creation of new functionally unique structures that self-assemble without needing external energetic inputs \cite{tanaka2006physical}.
Thermodynamic management of the environment further facilitates fine-grained control of the rate of self-propagation.
However, despite these affordances, as far as we know, an engineering testbed that mimics all aspects of prions, including their ability to exist in two distinct conformations, self-assemble into the prion state, and self-replicate like true biological prions, has yet to be developed.
We address this knowledge gap directly through this work.

In addition to its potential practical utility, our work offers several directions for further theoretical exploration.
While we limit our attention to two-dimensional frames, three-dimensional mechanical structures can be discovered and examined similarly using the techniques we describe here.
Symmetries can be imposed on the structures and subsequently utilized as foundational elements for constructing more complex physical or dynamical structures \cite{romano2020designing, ouellet2024breaking}.
Future work could also seek to reduce the degree of internal nodes, thereby reducing the potential for obstructions and easing the fabrication of useful macroscale mechanical prions. 
 
\newpage
\section{Acknowledgements}
We acknowledge the support of the Natural Sciences and Engineering Research Council of Canada (NSERC), the National Science Foundation through the University of Pennsylvania Materials Research Science and Engineering Center (MRSEC) (DMR-1720530), the National Science Foundation (IIS-1926757, DMR-1420530), the Paul G. Allen Family Foundation, and the Army Research Office (W911NF-16-1-0474, W011MF-191-244). 

\newpage

\bibliography{apssamp}

\newpage

\end{document}


\title[Supplementary Information]{Supplementary Materials}
 
\affiliation{
Department of Electrical \& Systems Engineering, University of Pennsylvania, Philadelphia, PA 19104 USA
}

\author{Dani S. Bassett} 
\affiliation{
Department of Bioengineering, University of Pennsylvania, Philadelphia, PA 19104 USA
}
\affiliation{
Department of Physics and Astronomy, University of Pennsylvania, Philadelphia, PA 19104 USA
}
\affiliation{
Department of Electrical \& Systems Engineering, University of Pennsylvania, Philadelphia, PA 19104 USA
}
\affiliation{
Department of Neurology, University of Pennsylvania, Philadelphia, PA 19104 USA
}
\affiliation{
Department of Psychiatry, University of Pennsylvania, Philadelphia, PA 19104 USA
}
\affiliation{
Santa Fe Institute, Santa Fe, NM 87501 USA
}
\affiliation{
Montreal Neurological Institute, McGill University, Montreal, QC H3A 2B4, Canada
}

\author{Lee C. Bassett}
\affiliation{
Department of Electrical \& Systems Engineering, University of Pennsylvania, Philadelphia, PA 19104 USA
}

\author{Kieran A. Murphy}
\affiliation{
Department of Bioengineering, University of Pennsylvania, Philadelphia, PA 19104 USA
}

\author{Shubhankar P. Patankar}
\affiliation{
Department of Bioengineering, University of Pennsylvania, Philadelphia, PA 19104 USA
}

\maketitle

Here, we provide a detailed account of our methods.
First, we outline the process of identifying mechanical structures capable of exhibiting prionic behavior.
Second, we describe how we identify the minimum energy path between two distinct conformations. 
Third, we describe the simulation process of validating prionic properties in a thermal bath. 
Fourth, we discuss the role of external and internal nodes in mechanical prions.
Finally, we offer design and experimental details for a macroscopic prion prototype.

\section{Designing mechanical prions}

In this section, we detail the process of searching for mechanical prions in the space of mechanical networks.
We begin by considering the external shape of potential prions and ``healthy'' mechanical structures.

This study employs a bar-joint linkage model to describe the structures.
Each structure, defined as a connected bar joint linkage, consists of a set $V$ of $N$ nodes and a set of edges $E$. 
The nodes $V = \left\lbrace v_1, v_2, \dots, v_N \right\rbrace \in \mathbb{R}^{N \times d}$, define the position of each node $v_i$ relative to the positions of nodes $v_1$ and $v_2$.
For clarity and readability, the pair $(V, E)$ is referred to simply as $V$ when the context makes it evident.
The model partitions the nodes into two distinct groups,  $V^{\text{Ex}}$ and $V^{\text{In}}$,  where $V^{\text{Ex}}$ represents exterior nodes enabling polymer binding, and $V^{\text{In}}$ represents interior nodes providing structural stability.
Thus, the model satisfies $V = V^{\text{Ex}} \cup V^{\text{In}}$ and $V^{\text{Ex}} \cap V^{\text{In}} = \varnothing$, ensuring that the two groups are mutually exclusive.
This framework offers a simple representation of the structures, enabling optimization for the specific task.

The structure $V$ represents the undeformed, idealized configuration, defined up to a Euclidean transformation. 
Most cases require examining the structure in its deformed or Euclidean-transformed state, referred to as $\bar{V}$.
To describe Euclidean transformations without deformation, the notation $\bar{V}(\mathbf{R}, \mathbf{t})$  expresses this relationship: $\bar{V}(\mathbf{R}, \mathbf{t}) = \left\lbrace \mathbf{R}v_i + \mathbf{t} \right\rbrace_{i=1}^N$, where  $\mathbf{R}$ defines a rotation matrix, and $\mathbf{t} \in \mathbb{R}^d$ represents a translation vector.
This framework focuses on the specific case of $d=2$, simplifying its application to two-dimensional structures.

The design of edges connecting the nodes plays a key role in defining the structure's geometry and functionality.
The set of edges $E$ consists of pairs  $(i, j)$, where each edge has a rest length $r_{ij}$.
This set partitions into two groups based on whether the edges form the outer shell of the structure:  $E^{\text{Ex}} = \left\lbrace (i, j) \mid (i, j) \in E, v_i, v_j \in V^{\text{Ex}} \right\rbrace$, and $E^{\text{In}} = \left\lbrace (i, j) \mid (i, j) \in E, v_i \in V^{\text{Ex}}, v_j \in V^{\text{Ix}} \right\rbrace$.
Here, $E^{\text{Ex}}$ includes the edges connecting external nodes, forming the outer perimeter of the structure, referred to as the external polygon, while $E^{\text{In}}$ includes the internal edges that provide structural stiffness. 
No edges connect two nodes from the set $V^{\text{Ix}}$.
This classification ensures a clear distinction between structural components, facilitating geometric and mechanical properties analyses.

The design and interaction of structures rely on simplified physical models that capture essential behaviors.
This framework models interactions between linked nodes using a harmonic potential with energy $U_{\text{bond}}(\bar{V}) = K \sum_{(i,j) \in E} \left( d(\bar{v}_i, \bar{v}_j) - r_{ij} \right)^2$ \cite{procyk2021coarse, rocklin2018folding}.
At rest and under a Euclidean transformation, $U_{\text{bond}}(V) = 0$ and $U_{\text{bond}}(\bar{V}(\mathbf{R}, \mathbf{t})) = 0$ because there is no deformation, i.e., $d(\bar{v}_i, \bar{v}j) = r_{ij}$.
Each node includes a Lennard-Jones (LJ) 12-6 potential to enable interactions between nodes across prions \cite{wang2020lennard}. 
The interaction energy of a structure $\bar{V}^i$ interacting with an ensemble of structures $S$ is given by 
$U_{\text{int}}(\bar{V}^i, S \setminus \bar{V}^i) = 2\epsilon \sum_{m \in S \setminus \bar{V}^i} \sum_{k=1}^N \sum_{l=1}^M \left( \frac{\sigma}{d(\bar{v}^{i}_k, \bar{v}^{m}_l)} \right)^{12} - \left( \frac{\sigma}{d(\bar{v}^{i}_k, \bar{v}^{m}_l)} \right)^{6}$
where $d(\bar{v}^{i}_k, \bar{v}^{m}_l)$ represents the distance between nodes $k$ and $l$ of structures $i$ and $m$, respectively.
This minimal model demonstrates that structures can be designed to exhibit prion-like behavior, providing a foundation for exploring their interactions and dynamics.

\subsection{External shape}

\begin{figure*}
\centering
  \includegraphics[width=0.8\textwidth]{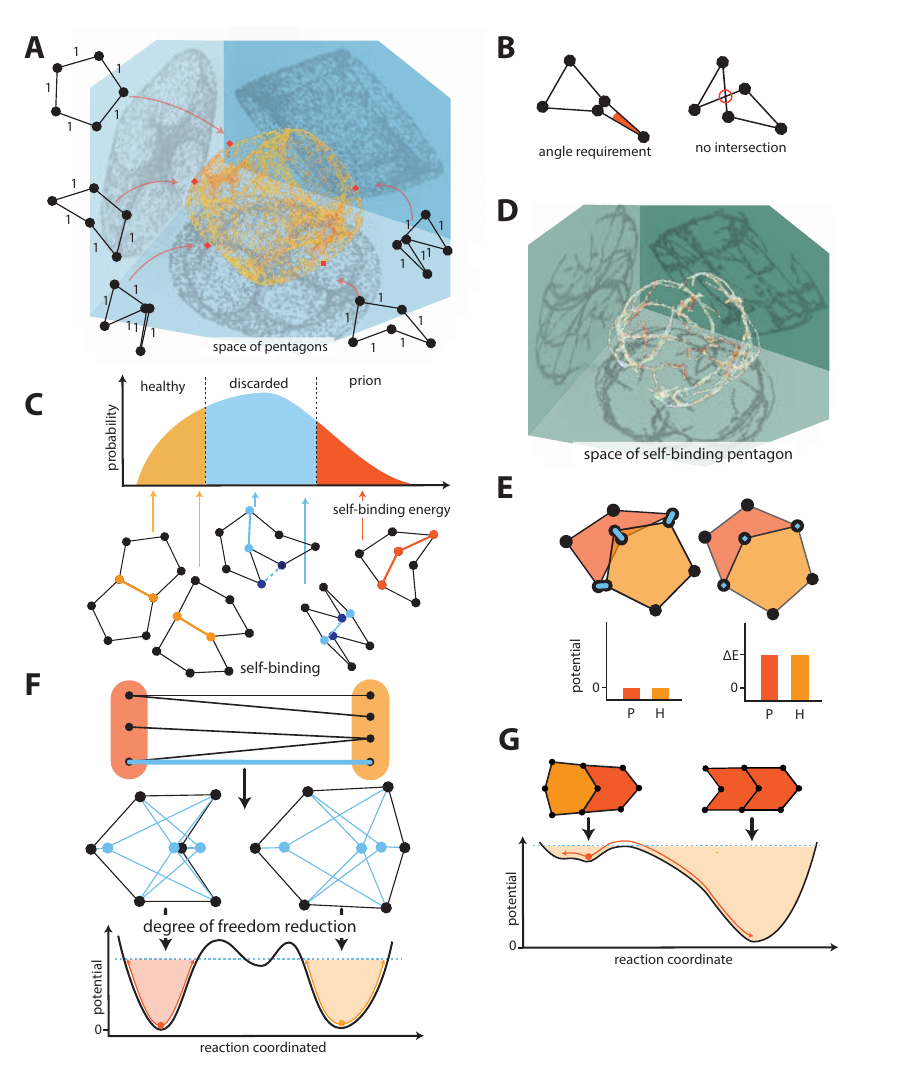}
  \caption{\textbf{Space exploration and prion design.}  
  (\textbf{A}) Pentagon space. Each point denotes a 2D pentagon structure. This space is topologically equivalent to a four-torus.  Shadows represent projections on three axes.
  (\textbf{B}) Polygon criteria. External polygons must meet two primary conditions: a specified minimal angle and no self-intersections.
  (\textbf{C}) Self-binding criteria. Polygons vary in their capabilities. Low-fitness polygons can primarily interact through pairs of nodes with themselves, whereas high-fitness polygons exhibit more interacting nodes, resulting in enhanced bound stability.
  (\textbf{D}) Filtered pentagon space. Polygons in this set adhere to the criteria outlined in panel B and are color-coded consistent with panel C. The space now appears fragmented, with prions frequently located at the ends of elongated filaments.
  }
  \label{fig:exploration}
\end{figure*}

We design the external shape in $d=2$ as a polygon with five edges, not constrained to a perfect pentagon ($|E^{\text{Ex}}|=5$).
We set the external edge length $r_{ij}=1$.
Two distinct structures drive the desired dynamics: the healthy structure (H) and the prionic structure (P).
Designing the simplest structure with prionic properties reveals that an external polygon of five nodes suffices. 
Structures with fewer nodes lack the necessary complexity to support conformal motions and binding. 
Consequently, this study explores mechanical prions within the regular pentagon space, which forms a 4-torus (see Figure \ref{fig:exploration}.A.) \cite{shimamoto2005spaces}.

This analysis focuses on generating and filtering pentagons to ensure they are constructible and adhere to the edge length criteria.
Pentagons are generated randomly by selecting three arbitrary angles in the range $[0,2\pi]$, with the remaining two angles determined based on the first three.
The set of random polygons is filtered to ensure that the distance between the starting and ending points is below $0.002$, satisfying the length criteria. 
Additional conditions ensure the polygons are simple (non-intersecting) and maintain a minimum internal angle of $10$ degrees, as shown in Figure \ref{fig:exploration}B (left).
This constraint on the minimum internal angle prevents external nodes from being too close to each other, reducing excessive interactions and preserving structural integrity.
These filtering steps create a set of pentagons that respect the length criteria and are realizable.

A metric is developed to evaluate the interaction properties of polygons and identify suitable structures for H and P conformations.
This metric investigate how the two structures, $(V^{\text{Ex}}_\text{H}, E^{\text{Ex}}_\text{H})$ and $(V^{\text{Ex}}_\text{P}, E^{\text{Ex}}_\text{P})$, corresponding to the H and P conformations, interact.
To evaluate potential interactions between polygons, the binding affinity between two structures is calculated using the following fitness value:
\begin{align}
    g(\bar{V}^{v\text{Ex}}_1(\mathbf{R}_1, \mathbf{t}_1), \bar{V}^{\text{Ex}}_2(\mathbf{R}_2, \mathbf{t}_2)) &= \frac{g_{\text{min}}^2}{2}\sum_{i\in \bar{V}^{\text{Ex}}_1 }\sum_{j \in \bar{V}^{\text{Ex}}_2} \max \left( \frac{1}{d(\bar{v}_{1,i}, \bar{v}_{2,j})^2}, g_{\text{min}}\right)\\
    f(V^{\text{Ex}}_1, V^{\text{Ex}}_2) &= \min_{ \mathbf{R}_1, \mathbf{t}_1, \mathbf{R}_2, \mathbf{t}_2} g(\bar{V}^{\text{Ex}}_1(\mathbf{R}_1, \mathbf{t}_1), \bar{V}^{\text{Ex}}_2(\mathbf{R}_2, \mathbf{t}_2)),
\end{align}
where  $g_{\text{min}}$ cap the diverging $1/r^2$ contribution. 
The value $g_{\text{min}}$ is set as $f_{\text{min}} = \frac{1}{d_{\text{min}}^2}$, with $d_{\text{min}} = 0.001\ell$. 
The Iterative Closest Points (ICP) shape registration algorithm minimizes this fitness for pairs of polygons \cite{rusinkiewicz2001efficient}. 
This metric provides a foundation for analyzing and comparing polygon interactions in prion-like systems.

We establish criteria to separate P and H structures based on their self-binding energy and interaction properties.
The P structure must self-bind with significantly lower energy than the H structure, enforced by filtering pairs of polygons to satisfy $f(V^{\text{Ex}}_\text{P}, V^{\text{Ex}}_\text{P}) \ll f(V^{\text{Ex}}_\text{H}, V^{\text{Ex}}_\text{H})$.
Figure \ref{fig:exploration}.C displays the various self-binding levels and their respective fitness values.
In the case of pentagons, this condition is met by identifying a P polygon ($V^{\text{Ex}}_\text{P}$), where three external nodes align with three nodes of another P polygon.
For healthy polygons, interactions between two copies are limited to two nodes, reducing their binding strength.
Additional filtering identifies pairs of polygons capable of self-assembling into tower-like structures of P polygons through a stacking mechanism resembling bowl stacking. 
Each polygon is classified as a potential healthy structure if its fitness $f$ is less than or equal to $n_{H}$ and as a potential prion if its fitness is greater than or equal to $n_{P}$.
For pentagons, we set $n_{H}$ and $n_{P}$ to 2.1 and 2.8, respectively. 
Figure \ref{fig:exploration}.D shows the refined space of polygons, highlighting self-binding capabilities using the color scheme from panel C.
These criteria allow us to evaluate polygon interactions and distinguish between potential H and P structures.

Having established the self-assembly properties of the P and H polygons themselves, the focus now turns to how these structures interact with each other.
The two complementary structures (H and P) must exhibit precise conformational matching to facilitate specific and stable binding \cite{woods2022shape}. 
However, a perfect fit would prevent the structures from leaving their respective energy minima during interaction.
A slight deformation is necessary to initiate the conformation change of the H polygon to bind with the P polygon. 
To achieve this, the filtering process identifies P and H pairs that interact imperfectly via three external nodes, with the H polygon requiring minimal deformation for binding.
This imperfection helps disrupt the rate symmetry between free and bonded states.
The criteria enforce $f(V^{\text{Ex}}_\text{H}, V^{\text{Ex}}_\text{H}) < f(V^{\text{Ex}}_\text{H}, V^{\text{Ex}}_\text{P}) < f(V^{\text{Ex}}_\text{P}, V^{\text{Ex}}_\text{P})$, ensuring that the binding strength between P and H is greater than that of the healthy dimer but lower than that of the prionic dimer.
These criteria let us identify P and H polygons that meet the necessary interactions to lead to prion dynamics.

\addtocounter{figure}{-1}
\begin{figure} [t!]
  \caption{(Previous page.) (\textbf{E}) Boundary Mismatch. The mechanism necessitates a mismatch in the interacting boundaries of the two polygons. This discrepancy elevates the potential energy and facilitates a shift in the pathway.
  (\textbf{F}) Internal Nodes. Pairs are derived from the sets of healthy polygons and prions. Internal nodes and edges (depicted in blue) are positioned to ensure zero potential energy for both external conformations. A potential energy barrier separates these two conformations.
  (\textbf{G}) Prionic Behavior. The search procedure identifies structures that, similar to prions, prefer a specific conformation during interactions.}
\end{figure}


\subsection{Internal nodes}

The candidate mechanical prions require both flexibility for conformational changes and rigidity to preserve their shape in a heat bath. 
The selected pairs of pentagons exhibit multiple conformational degrees of freedom, but they must maintain sufficient rigidity to remain intact in a bath. 
To achieve this, internal nodes ($V^{\text{In}}$) are introduced and connected to the exterior structure designed earlier, ensuring mechanical stability \cite{chen2022modular}.
Both structures share the same nodes, edges, and connections, enabling transformation between them through purely conformal changes.

Ensuring rigidity in mechanical prions requires careful selection of the number of internal nodes and their connecting edges.  
The Maxwell-Calladine index theorem \cite{rocks2024integrating} state that rigidity happens when $\left| E^{\text{Ex}}\right| = 2  \left| V^{\text{In}} \right| +  \left| E^{\text{In}} \right| - 3$.
When this equation is met, only rigid-body transformations solve the energy equation  $\frac{\partial U_{\text{bond}}}{\partial \mathbf{v_i}} = 0$ \cite{chen2022modular}.
Solutions to the Maxwell-Calladine equations are generally not unique. 
Some configurations significantly restrict the system’s dynamics, as seen when  $\left| E^{\text{Ex}}\right| = 5$ and $\left| V^{\text{In}} \right| = 1$ in the case of pentagonal external polygon.
To balance rigidity and dynamics, the configuration with $\left| V^{\text{In}} \right| = 2$ is selected, distributing the edges evenly between the two internal nodes.
This configuration ensures sufficient rigidity while preserving the desired flexibility for functional dynamics.

The Maxwell-Calladine equation does not provide information on the node positions and edge lengths, $ r_{ij} $, which must be determined.
Given the external shapes of the healthy and prionic particles, $V^{\text{Ex}}_{\text{H}}$ and $V^{\text{Ex}}_{\text{P}}$, our goal is to determine $V^{\text{In}}_{\text{H}}$ and $V^{\text{In}}_{\text{P}}$, such that,
\begin{align}
    d(v_i\in V^{\text{Ex}}_{\text{H}}, v_j \in V^{\text{In}}_{\text{H}}) = d(v_i\in V^{\text{Ex}}_{\text{P}}, v_j \in V^{\text{In}}_{\text{P}}) \qquad \forall (i,j)\in E^{\text{In}}.
    \label{eqn:dist}
\end{align}
A conformational solver \cite{ kim2019conformational,kim2022nonlinear} solves this set of equations, ensuring that every edge in $E^{\text{In}}$ maintains a consistent length in both the healthy and prionic structures.
For each external pair of shapes identified in the previous section, we construct an ensemble of potential internal nodes that solve equation \ref{eqn:dist}.
The energetic properties of these pairs must then be evaluated to determine if they exhibit prion-like behavior.
This approach ensures compatibility between the two structures, allowing them to be stable in both conformations.

We evaluate solutions for internal node configurations to ensure complete coverage of the solution manifold. 
We test $10,000$ solutions for each pair with a length discrepancy tolerance of $0.01\ell$, discarding duplicate solutions.
We then sample the identified manifold to ensure complete coverage.
In this model, the solutions remain independent because the internal nodes do not interact with each other through edges. 
As a result, the complete manifold is the product of the manifolds for each individual internal node.
Figure \ref{fig:internal} illustrates such a manifold.
Determining the positions of the internal nodes and their connection topology establishes the lengths of all edges $r_{ij}$, generating pairs of complete H and P structures.

\section{Energy landscape}

From the previous section, we selected the positions of external and internal nodes and defined their connections to ensure the existence of two zero-energy conformations, their ability to interconvert, and proper self-assembly.
This section aims to identify structure pairs from those structures where the interaction with the prionic conformation reduces the activation energy barrier between those two conformations.
This mechanism enables prionic propagation and creates a rate imbalance favoring the H-to-P transition in the presence of P.

We use the Nudged Elastic Band (NEB) method \cite{henkelman2000climbing} in LAMMPS \cite{dias2021molecular} to efficiently identify transition paths and energy barriers between molecular conformations, laying the foundation for prion screening. 
This method connects replicas of the structure with virtual elastic bands, with the chain's two extremities fixed in distinct conformations while minimizing the system’s energy.
The approach determines the minimum energy path, where the highest energy peak along this path represents the activation energy.
The H conformation (H-P dimer) serves as the initial configuration, and the P conformation (P-P dimer) represents the final configuration, allowing us to evaluate the energy barrier in the free (bonded) cases. 
An initial relaxation step ensures the newly associated structures will settle into their nearest energy minimum.
For both cases (bonded and free), we use a step size of 0.001 with 12 replicas.
Although NEB provides a static representation of the energy landscape and excludes kinetic and entropic considerations, it operates several orders of magnitude faster than extended dynamic simulations. 
This efficiency makes it a valuable tool for screening potential prions. 
By combining NEB with dynamical analysis, we ensure a comprehensive evaluation of candidate prion structures, addressing the limitations of static methods.

\section{Dynamical Analysis}

\begin{figure*}
\centering
  \includegraphics[width=0.95\textwidth]{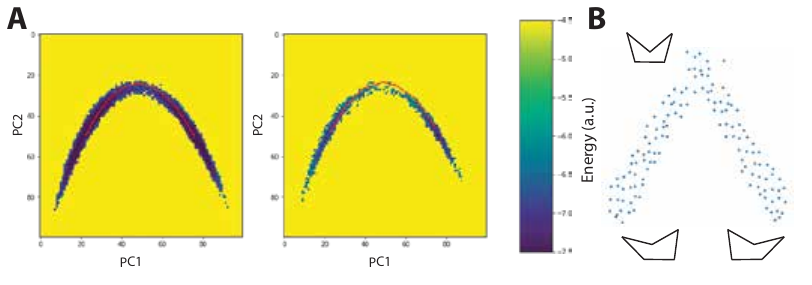}
  \caption{\textbf{The system's properties.}  
  (\textbf{A}) The energy landscape is attainable from the two conformations (healthy and prion) at the temperature under analysis. 
  Conformations are projected onto the 2D plane using principal components analysis (PCA).
  The red trajectory indicates the minimum energy path between the two conformations in the free case.   
  (\textbf{B}) Illustration of the external shape manifold used to study its impact on prionic properties.
  }
  \label{fig:manifold}
\end{figure*}

Through this search process, we discovered several candidate mechanical prions in the space of pentagonal bar-joint linkages. 
Although their static energy landscape suggests prionic behavior, it remains essential to evaluate them under dynamic conditions, wherein multiple effects can appear that prevent the transition from P-H to P-P. 
Examples of these effects include internal node binding issues, other stable configurations, and entropy inhibiting the conformational changes \cite{tzeng2012protein}.
Hence, we must evaluate the behavior for the specific pair of configuration states under Langevin dynamics.

We model the dynamics of conformational changes using the Langevin formalism implemented in the Large-scale Atomic/Molecular Massively Parallel Simulator (LAMMPS).
The force acting on a node $i$ follows the equation:
\begin{align*}    
    F^{i}_k(\Bar{V}_k^i, S\setminus \Bar{V}^i) &= - \nabla U(\Bar{V}_k^i, S\setminus \Bar{V}^i) \\ 
    & \qquad - \gamma*m*\dot{\Bar{V}}_k^i + F_{k,\text{solvent}}^{i},
\end{align*}
where $\nabla U(\Bar{V}_k^i, S\setminus \Bar{V}^i)$ represents the gradient of the system's potential energy.
The second term, $\gamma*m*\dot{\Bar{V}}_k^i$, describes the frictional drag and/or viscous damping.
For simplicity, we normalize the damping coefficient $\gamma$ to $1$. 
The third term, $F_{k,\text{solvent}}^{i}$, accounts for the force exerted by an imaginary solvent particle, which exchanges energy with the nodes through random interactions and is proportional to $2 \gamma T m $ \cite{dias2021molecular}. 
Although slower than the NEB method, this approach provides a more nuanced and dynamic picture of conformational changes, improving upon the static representation offered by NEB.

We investigate conformational dynamics by quantifying structural deformations, modeling transitions with a Markovian approach, and tracking population changes in interacting and non-interacting states.
Structural deformations are measured as the sum of squared differences in their internal angles relative to reference conformations, and each structure is labeled as the closest reference conformation, either P or H. 
These labels facilitate the dynamical analysis through a discrete-time Markovian model, which derives transition rates via model fitting over a coarse-grained time scale.
We investigate the evolution of an ensemble comprising P and H structures and track the concentrations of both interacting (bound P-H and P-P) and non-interacting (free P and H) structures. 
This combined approach allows us to measure how interactions shape the system's behavior over time.

We investigate the dynamical stability and interaction behavior of candidate pairs with promising energy landscapes across varying temperatures ranging from $1e^{-8} \frac{1}{\epsilon}$ to $1e^{-3} \frac{1}{\epsilon}$.
Each simulation runs for  $12,000$ time points, recording the positions of all structures at $600$ evenly spaced intervals.
The number of time steps doubles for each failed run until the simulations converge.
Simulations start in one of three configurations, each containing 30 structures within a $15 \times 15$ box.
The first two configurations (H or P) evaluate the stability of healthy and prionic structures independently, while the third configuration (H-P) examines pairs of prionic and healthy structures. 
This systematic approach provides insights into the stability and transition rates of the candidate structures under varying conditions.

We establish simulation conditions to ensure accurate and controlled modeling of molecular dynamics. 
Fixed boundary conditions are applied along the x- and y-axes, constraining molecules to the same z-plane. 
A neighbor list is constructed using a simple nearest-neighbor search algorithm with a skin distance of 1 unit. 
Lennard-Jones interactions are employed with a cutoff distance of 2.5 units.
The simulation maintains a constant number of particles, volume, and energy integration for all masses. 
A Langevin thermostat is applied to all masses with a damping coefficient of 1. 
The LAMMPS minimization procedure minimizes the initial structure with a convergence tolerance of 0.000001 for energy and forces. 
These conditions ensure stability and precision in the molecular dynamics simulations.

\subsection{Analyzing trajectories of conformational change}

In the manuscript, we examine \textit{in silico} trajectories of conformational change.
We assign labels to intermediate states during simulations and analyze the resulting coarse-grained sequences assuming the Markov property.
The labels denote one of three general states: healthy, prionic, or denatured.
To assign these labels, we compute distances between each intermediate state and the prionic and healthy configurations. 
For a given structure in conformation $c$, let $\Delta$ be the set of all triplets of connected nodes $(i,j,k)$, and let $\theta^c_{ijk}$ be the angle between them.
We define the distance $D$ as
\begin{align*}    
    D(\Bar{V}^n, \Bar{V}^\text{ref}) = \sum_{(ijk)\in \Delta} (\theta^{\text{ref}}_{ijk} - \theta^{n}_{ijk})^2.
\end{align*}
We first measure the distance between the healthy and prionic states, $D_{max} = D(V_P,V_H)$.
For structure $k$, we then measure distances to the healthy and prionic configurations as $D_H = D(\Bar{V}^k,V_H)$ and $D_P = D(\Bar{V}^k,V_P)$, respectively. 
We assign the label H to structure $k$ if $D_H < D_P$; otherwise, we assign the label P.

We use principal component analysis (PCA) to facilitate the visualization of conformational dynamics across all temperatures.
This method projects states at all temperatures onto the nearest point on the reaction pathway. 
The transformation is learned using $\theta^c_{ijk}$ as row vectors into the data matrix containing all time points and applying PCA, defining reaction coordinates as the first principal component.
This process is carried out dynamically across our extensive spectrum of analyzed temperatures, all within the same matrix, to capture the complete range of motion.
This component accounts for more than 95\% of the data variance (see Figure \ref{fig:manifold}.A).
While highly effective, this approach requires extensive numerical simulations to obtain sufficient data for PCA at all temperatures. 
This ensures a comprehensive representation of the system's dynamics.

\subsection{Rate computation}

We compute reaction rates and steady-state ratios from coarse-grained sequences of conformational states modeled as Markov processes.
Rate curves are adapted to fit a modified Arrhenius equation, as described in \cite{kohout2021modified}, of the form:
\begin{align*}    
    k = k_\infty \exp \left[ -\frac{E_A}{R(T-T_0)} \right].
\end{align*}
where $ k_\infty $ represents the maximum rate constant, $ E_A $ is the activation energy, $ R $ is the universal gas constant, and $ T_0 $ is the temperature offset. 
This equation accounts for deviations from classical Arrhenius behavior, particularly at low temperatures, which are relevant to prionic dynamics.

\begin{figure*}
\centering
\includegraphics[width=0.7\textwidth]{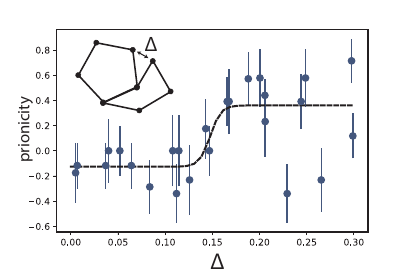}
  \caption{\textbf{Role of external nodes.} The prionic behavior is examined as a function of the mismatch distance, $\Delta$, between the pair of prions and healthy conformations. Each pair features similar internal node positions and minor differences in external geometries compared to the example provided in the paper. The dashed line represents a sigmoid fit to the data points. 
  }
  \label{fig:delta}
\end{figure*}

\section{Additional considerations}

\subsection{Properties of the External Node}
In this section, we describe how we analyze the impact on the prionic property of the external node. 
The goal is to study slightly deformed polygons with similar internal structures and assess the effects of such deformations.
To achieve this goal, we begin with the initial pairs from the main text and choose polygons close to the prionic external polygon within the filtered manifold (see Figure \ref{fig:exploration}.D).
The manifold segment resembles an arrowhead, as shown in Figure \ref{fig:manifold}, where the vertical axis in the illustration corresponds to the base angle.
At the top, the symmetric case is observed, and mirror reflection along the vertical axis appears on either side of the arrowhead.
The dimension corresponding to the arrow's thickness is associated with the cusp angle.
At this stage, only the external polygon is present. 
Internal nodes are introduced to minimize the distance to the original prion. 
This aim is achieved by navigating the manifold formed by the internal points and selecting the nearest one.

The difference, $\Delta$, is determined by the initial distance between nodes positioned for binding before the actual interaction.
A higher value suggests a lesser fit between the two shapes, implying increased stress during their interaction.
The difference in activation temperature, called here prionicity, is defined as 
\begin{equation}
    \text{Pr} = \frac{T_{free} - T_{bound}}{T_{free} + T_{bound}},
\end{equation}
where $T_{free}$ and $T_{bound}$ represent the minimum temperatures at which over $40\%$ of the limit distribution consists of prions in the free and bound scenarios, respectively.
The $40\%$ threshold is selected as it is below the typical $50\%$ (even mixture) observed for the free case.

In Figure \ref{fig:delta}, we plot prionicity, $\text{Pr}$, as a function of the mismatch, $\Delta$, between the key-lock areas for various mechanical prions.
All prions share nearly identical internal and external shapes, with the exception of the external lock region.
We measure $\Delta$ as the smallest distance at which nodes from the key and the lock can be paired in P-H dimers without forcing conformational changes. 
We find that prionicity increases with increasing key-lock mismatch. 

\subsection{Properties of the Internal Nodes}
In this section, we examine the influence of internal nodes on the prionic behavior of mechanical structures.
We begin by detailing some of their characteristics and then provide insights into the methodology employed to reach these conclusions.

\begin{figure*}
\centering
  \includegraphics[width=0.98\textwidth]{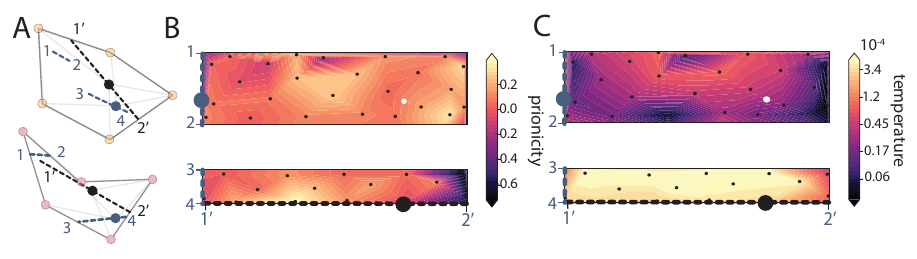}
  \caption{\textbf{Role of internal nodes}  
 (\textbf{A}) The prionicity, a metric used to quantify the reduction in transition temperature caused by the interaction of the dimer compared to that of the individual monomers, as a function of the position of the two internal nodes.
  (\textbf{B}) The transition temperature of the prion.  (\textbf{C}) The prionicity as a function of the mismatch distance $\Delta$ for a pair of prions and healthy conformations with distinct internal nodes and minimal variations in their external shapes.
  }
  \label{fig:internal}
\end{figure*}

As mentioned earlier, an important property of the placement of the internal node in our framework is that the lengths of the edges between the internal and external nodes must be the same in both conformations to meet the system's energy requirements. 
Therefore, given the external shape, the number of internal nodes, and their connections to the external nodes, the internal nodes can only occupy certain positions within the polygon.
Figure \ref{fig:internal}.A displays the potential positions that the internal node can assume for the primary structure discussed in the paper.
The internal node's position in the healthy structure dictates its location in the prion and \emph{vice versa}.
Furthermore, the positions of the individual internal nodes are independent of one another.
In this particular polygon, with its connection pattern to the external node, the potential placement space for internal nodes is characterized by near-linear trajectories. 
Specifically, two separate connected components define the placement set for one node. 
While these sets are not always strictly linear, they predominantly tend to be.
Given that no edges are connecting them and the external polygon is fixed \emph{a priori}, the two systems of equations operate independently.
Therefore, the solution space for both nodes in our system is the product of the solution space for each node. 
Consequently, the design of our example prion has two additional degrees of freedom, which affect its stability and transition temperature.

The system's prionicity exhibits intricate variations based on the internal node's position, yet it maintains a similar magnitude between the two connected components of the domain (see Figure \ref{fig:internal}.B).
The white point indicates the position of the structures shown in the paper.
Given its closeness to a region with reduced prionicity and more favorable zones, this placement might appear counterintuitive.
However, this choice was driven by two primary factors. 
First, it was chosen because its transition temperature was within our desired range: not so high as to complicate convergence in large-scale numerical simulation, yet not so low as to render the physical model unstable (see Figure \ref{fig:internal}.C).
Second, to construct the physical model, the central nodes must not interfere with one another and remain sufficiently distant throughout the standard range of motion to avoid interaction. 
While the present selection can be further optimized, it sufficiently demonstrates the effect while potentially leaving room for enhancement.
Interestingly, most positions of the interior nodes give rise to valid prions that perform adequately, implying that precise placement of the interior nodes is less critical.

The relation between prionicity and the transition temperature is quite complex, as the internal node placement affects $T_{free}$ and $T_{bound}$ differently.
The temperatures $T_{free}$ are shown in Figure \ref{fig:internal}.C, where we observe a stark contrast between the two disconnected regions of the solution space. 
$ T_{free} $ also indicates that the two internal nodes engage in synergic interactions.
Thus, even though their placements, based solely on energy levels, appear independent, the final selection must consider their interactions.

\begin{figure*}
\centering
  \includegraphics[width=0.95\textwidth]{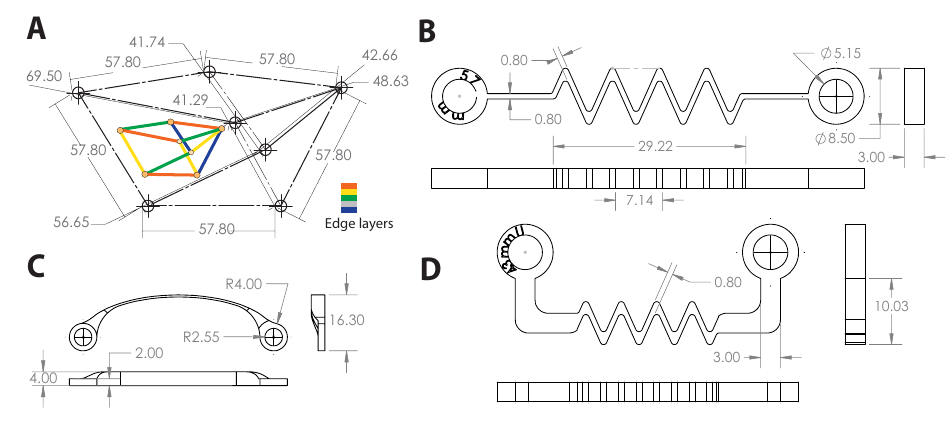}
  \caption{\textbf{Macroscopic model drawing.}  
  (\textbf{A}) Edge dimensions with color-differentiated layers shown in the inset in millimeters. The lengths of these edges closely align with the structure presented in the paper. 
  Any variations in length may be attributed to the precision level maintained during the construction phase. The arrangement prevents overlaps and collisions.
  (\textbf{B}) The primary edge type in the model.
  The central zigzag functions as a rigid spring, with force modulated by the zigzag's thickness (0.8 mm) and the edge's thickness (3 mm).
  (\textbf{C}) The secondary edge type, utilized only once in the actual model, offers a broader range of expansion and contraction. 
  While the entire model could be constructed using this edge type, it would compromise the desired aesthetic appeal.
  (\textbf{D}) For the primary edge type, occasional translations or rotations of the spring component are necessary to prevent collisions.}
  \label{fig:drawing}
\end{figure*}

\section{Experimental Demonstration}

This section provides insights into constructing the macroscopic model and elaborates on the experimental setup employed.
We created a macroscopic model for proof of concept, featuring external edges of 57.80 mm in length (see Figure \ref{fig:drawing}.A).
To accurately replicate our design, we needed to devise edges capable of extending and retracting based on a harmonic potential.
Fortunately, the model's performance is not critically dependent on the exact specifications of its components. 
We chose two types of edges that, while not perfectly matching the desired properties, still ensure the system's functionality.
The first type, similar to a spring, contracts readily but exhibits limited extension capabilities (see Figure \ref{fig:drawing}.B).
The second type, similar to a leaf spring, offers superior extension properties compared to the first but is less robust (see Figure \ref{fig:drawing}.C).
Either can serve as the primary edge type for the structure independently, but employing both proved effective.
In the physical realm, edges cannot pass through one another, which is a fundamental distinction from our simulation. This characteristic necessitates careful consideration and adaptation in our macroscopic model.
We employed edges modified with off-center springs for the central nodes to address this issue, as depicted in Figure \ref{fig:drawing}.D. 
Additionally, we arranged the edges over several layers, as illustrated in Figure \ref{fig:drawing}.A, incorporating an unused layer to accommodate the sagging of the central nodes.

\begin{figure*}
\centering
  \includegraphics[width=0.65\textwidth]{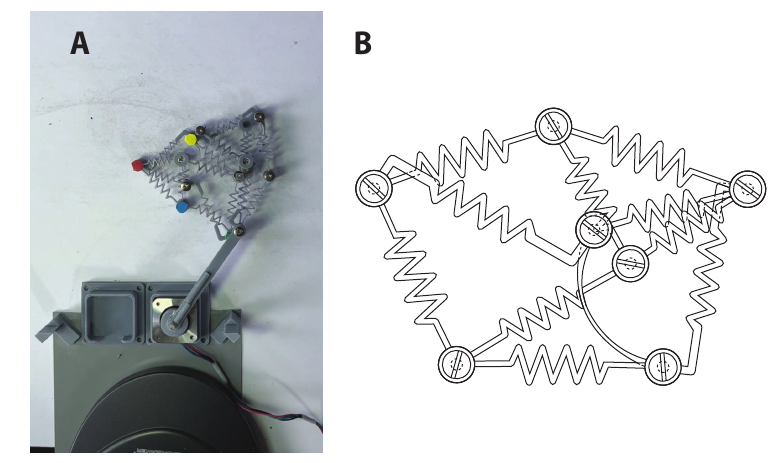}
  \caption{\textbf{Macroscopic model experiment.}  
  (\textbf{A}) Experimental setup featuring two interacting structures agitated by a stepper motor.
  (\textbf{B}) Schematic of the fully assembled macroscopic model used in the experiment. 
  }
  \label{fig:experiment}
\end{figure*}

All edges were 3d printed using ultraviolet-sensitive resin and assembled with Chicago screws, as depicted in Figure \ref{fig:experiment}.
3R mm rare earth magnets were affixed to the screw's bottom to allow interaction between the external nodes, and a printed cap was added for centering.
The cap's thickness was calibrated to regulate the holding force between nodes, thereby ensuring that structures remain bound when three nodes interact, but not when only two nodes interact.

In the experimental setup, an arm measuring 100mm in length is manipulated by a stepper motor.
The structures are attached to the arm via one of their nodes, leveraging the pre-existing magnet (see Figure \ref{fig:experiment}.C).
The stepper motor undergoes oscillations spanning 1 radian for 20 cycles in the experimental setup.
Between each oscillation set, the conformation is reset, and the angle of the structure around its attachment point to the arm is randomized.
In the experiment, the speed is regulated by the number of steps, each measuring 1.8 degrees, ranging from 50 to 250 steps per second.
Conformations were manually assessed, given the straightforward nature of the task and the macro structure's distinct transition to the prion conformation.
Due to the limited extension range of the edges, the macro model is constrained to transition only between the healthy and prion conformations.
Unlike the computational model, the physical version necessitates disassembly for denaturation and is susceptible to breakage under excessive force.

\bibliography{apssamp}